\documentclass[twocolumn,showpacs,aps,prl,superscriptaddress,floatfix]{revtex4}

\usepackage{graphicx}
\usepackage{dcolumn}
\usepackage{amsmath}
\usepackage{epsfig}
 
\RequirePackage{xspace}

\usepackage{relsize}
\def\babar{\mbox{\slshape B\kern-0.1em{\smaller A}\kern-0.1em
    B\kern-0.1em{\smaller A\kern-0.2em R}}}

\def\Kbar  {\kern 0.2em\overline{\kern -0.2em K}{}\xspace}

\def\Kz    {\ensuremath{K^0}\xspace}
\def\Kzb   {\ensuremath{\Kbar^0}\xspace}
\def\KzKzb {\ensuremath{\Kz \kern -0.16em \Kzb}\xspace}
\def\Kp    {\ensuremath{K^+}\xspace}
\def\Km    {\ensuremath{K^-}\xspace}

\def\KpKm  {\ensuremath{\Kp \kern -0.16em \Km}\xspace}
\def\KS    {\ensuremath{K^0_{\scriptscriptstyle S}}\xspace} 
\def\KL    {\ensuremath{K^0_{\scriptscriptstyle L}}\xspace} 
\def\Kstarz  {\ensuremath{K^{*0}}\xspace}
\def\Kstarzb {\ensuremath{\Kbar^{*0}}\xspace}
\def\Kstar   {\ensuremath{K^*}\xspace}

\def\Dbar    {\kern 0.2em\overline{\kern -0.2em D}{}\xspace}

\def\Dz      {\ensuremath{D^0}\xspace}
\def\Dzb     {\ensuremath{\Dbar^0}\xspace}
\def\DzDzb   {\ensuremath{\Dz {\kern -0.16em \Dzb}}\xspace}
\def\Dp      {\ensuremath{D^+}\xspace}
\def\Dm      {\ensuremath{D^-}\xspace}

\def\DpDm    {\ensuremath{\Dp {\kern -0.16em \Dm}}\xspace}

\def\B       {\ensuremath{B}\xspace}
\def\Bbar    {\kern 0.18em\overline{\kern -0.18em B}{}\xspace}

\def\Bz      {\ensuremath{B^0}\xspace}
\def\Bzb     {\ensuremath{\Bbar^0}\xspace}
\def\BzBzb   {\ensuremath{\Bz {\kern -0.16em \Bzb}}\xspace}
\def\Bu      {\ensuremath{B^+}\xspace}
\def\Bub     {\ensuremath{B^-}\xspace}

\def\BpBm    {\ensuremath{\Bu {\kern -0.16em \Bub}}\xspace}

\def\BorBbar    {\kern 0.18em\optbar{\kern -0.18em B}{}\xspace}
\def\DorDbar    {\kern 0.18em\optbar{\kern -0.18em D}{}\xspace}
\def\KorKbar    {\kern 0.18em\optbar{\kern -0.18em K}{}\xspace}

\def\jpsi     {\ensuremath{{J\mskip -3mu/\mskip -2mu\psi\mskip 2mu}}\xspace}

\mathchardef\Upsilon="7107
\def\Y#1S{\ensuremath{\Upsilon{(#1S)}}\xspace}% no space before {...}!

\def\FourS {\Y4S}

\mathchardef\Deltares="7101
\mathchardef\Xi="7104
\mathchardef\Lambda="7103
\mathchardef\Sigma="7106
\mathchardef\Omega="710A

\def\Deltabar{\kern 0.25em\overline{\kern -0.25em \Deltares}{}\xspace}
\def\Lbar{\kern 0.2em\overline{\kern -0.2em\Lambda\kern 0.05em}\kern-0.05em{}\xspace}
\def\Sigbar{\kern 0.2em\overline{\kern -0.2em \Sigma}{}\xspace}
\def\Xibar{\kern 0.2em\overline{\kern -0.2em \Xi}{}\xspace}
\def\Obar{\kern 0.2em\overline{\kern -0.2em \Omega}{}\xspace}
\def\Nbar{\kern 0.2em\overline{\kern -0.2em N}{}\xspace}
\def\Xb{\kern 0.2em\overline{\kern -0.2em X}{}\xspace}

\def\mes        {\mbox{$m_{\rm ES}$}\xspace}

\newcommand{\tev}{\ensuremath{\mathrm{\,Te\kern -0.1em V}}\xspace}
\newcommand{\gev}{\ensuremath{\mathrm{\,Ge\kern -0.1em V}}\xspace}
\newcommand{\mev}{\ensuremath{\mathrm{\,Me\kern -0.1em V}}\xspace}
\newcommand{\kev}{\ensuremath{\mathrm{\,ke\kern -0.1em V}}\xspace}
\newcommand{\ev}{\ensuremath{\mathrm{\,e\kern -0.1em V}}\xspace}
\newcommand{\gevc}{\ensuremath{{\mathrm{\,Ge\kern -0.1em V\!/}c}}\xspace}
\newcommand{\mevc}{\ensuremath{{\mathrm{\,Me\kern -0.1em V\!/}c}}\xspace}
\newcommand{\gevcc}{\ensuremath{{\mathrm{\,Ge\kern -0.1em V\!/}c^2}}\xspace}
\newcommand{\mevcc}{\ensuremath{{\mathrm{\,Me\kern -0.1em V\!/}c^2}}\xspace}
\def\invfb   {\ensuremath{\mbox{\,fb}^{-1}}\xspace}

\def\mus  {\ensuremath{\rm \,\mus}\xspace}

\def\ps   {\ensuremath{\rm \,ps}\xspace}

\def\mus        {\ensuremath{\,\mu{\rm s}}\xspace}    %% microsecond
\def\ps         {\ensuremath{{\rm \,ps}}\xspace}  %% picosecond

\def\to                 {\ensuremath{\rightarrow}\xspace}

\def\pep2{PEP-II}
\def\BF{$B$ Factory}

\def\gsim{{~\raise.15em\hbox{$>$}\kern-.85em
          \lower.35em\hbox{$\sim$}~}\xspace}
\def\lsim{{~\raise.15em\hbox{$<$}\kern-.85em
          \lower.35em\hbox{$\sim$}~}\xspace}

\def\eps{\varepsilon\xspace}

\def\CP                {\ensuremath{C\!P}\xspace}
\def\stwob{\ensuremath{\sin\! 2 \beta   }\xspace}

\def\mistag{\ensuremath{w}\xspace}

\def\deltaz{\ensuremath{{\rm \Delta}z}\xspace}
\def\deltat{\ensuremath{{\rm \Delta}t}\xspace}
\def\deltamd{\ensuremath{{\rm \Delta}m_d}\xspace}

\xspace

\newcommand{\jplBase}        {Phys.\ Lett.\xspace}

\newcommand{\npBase}         {Nucl.\ Phys.\xspace}
\newcommand{\zpBase}         {Z.\ Phys.\xspace}

\newcommand{\np}        [1]  {\npBase\ {\bf #1}}
\newcommand{\plb}       [1]  {\jplBase\ B~{\bf #1}}

\newcommand{\zpc}       [1]  {\zpBase\ C~{\bf #1}}

\def\jetset74   {\mbox{\tt Jetset \hspace{-0.5em}7.\hspace{-0.2em}4}\xspace}

\providecommand{\CKS}{\mbox{\ensuremath{C}}}
\providecommand{\CKSB}{\mbox{\ensuremath{\overline{C}}}}
\providecommand{\SKS}{\mbox{\ensuremath{S}}}
\providecommand{\SKSB}{\mbox{\ensuremath{\overline{S}}}}

\newcommand{\BABARPubYear}     {03}
\newcommand{\BABARPubNumber}  {016}
\newcommand{\SLACPubNumber} {10394}

\begin{document}
\noindent
\babar-PUB-\BABARPubYear/\BABARPubNumber\\
SLAC-PUB-\SLACPubNumber\vskip 0.4cm

\title{
\large \bf  Measurement of the Ratio of Decay Amplitudes for
  $\Bzb\to \jpsi\Kstarz$ and $\Bz\to \jpsi\Kstarz$}

%\input pubboard/authors_mar2004.tex
%% author list as of 02-Mar-2004 (602 authors)
%
\author{B.~Aubert}
\author{R.~Barate}
\author{D.~Boutigny}
\author{F.~Couderc}
\author{J.-M.~Gaillard}
\author{A.~Hicheur}
\author{Y.~Karyotakis}
\author{J.~P.~Lees}
\author{V.~Tisserand}
\author{A.~Zghiche}
\affiliation{Laboratoire de Physique des Particules, F-74941 Annecy-le-Vieux, France }
% ...
\collaboration{The \babar\ Collaboration}
\noaffiliation

\date{\today}% It is always \today, today, but you may specify any date with \date.

\begin{abstract}
We have measured the time-dependent decay rate for the process
$B\!\to\!\jpsi\Kstarz(892)$ in a sample of about 88 million
$\FourS\!\to\! B\Bbar$ decays collected with the \babar\ detector at the
\pep2\ asymmetric-energy \BF\ at SLAC. In this sample we study
flavor-tagged events in which one neutral $B$ meson is reconstructed in
the $\jpsi\Kstarz$ or  $\jpsi\Kstarzb$ final state. We measure the
coefficients of the cosine and sine terms in the time-dependent
asymmetries for $\jpsi\Kstarz$ and $\jpsi\Kstarzb$, find them to be
consistent with the Standard Model expectations, and set upper limits at
90\% C.L. on the decay amplitude ratios $|A
(\Bzb\!\to\!\jpsi\Kstarz)|/|A (\Bz\!\to\!\jpsi\Kstarz)| < 0.26$ and 
$|A (\Bz\!\to\!\jpsi\Kstarzb)|/|A (\Bzb\!\to\!\jpsi\Kstarzb)|
<0.32$. For a single ratio of wrong-flavor to favored amplitudes for \Bz
and \Bzb combined, we obtain an upper limit of $0.25$ at $90\%$ C.L.  
\end{abstract}

% PACS, the Physics and Astronomy Classification Scheme.

\maketitle

The Standard Model of electroweak interactions describes \CP\ violation 
in weak interactions of quarks by the presence of a complex phase in
the three-generation Cabibbo-Kobayashi-Maskawa (CKM) quark-mixing
matrix~\cite{CKM}. In this framework, the \CP  asymmetries in the
proper-time distributions of neutral $B$ decays to $\jpsi\KS$ and
$\jpsi\KL$ are directly related to the \CP-violation parameter
\stwob~\cite{BCP}. The time-dependent \CP asymmetries for $\jpsi\KS$ and  
$\jpsi\KL$ are of opposite sign and, to a very good approximation, equal
in magnitude~\cite{ligeti}.  The decay $\Bz\!\to\!\jpsi\KS$ 
($\Bz\!\to\!\jpsi\KL$) proceeds through the CKM-favored, color-suppressed
decay $\Bz \!\to\! \jpsi \Kz$~\cite{chargeconj} followed by $\Kz\!\to\!\KS$
($\Kz\!\to\!\KL$). The so-called wrong-flavor \Bz\ decay amplitude to the
opposite strangeness final state $\Bz \!\to\! \jpsi \Kzb$ is expected to be
negligible in the Standard Model~\cite{ligeti}. Interference between a
wrong-flavor amplitude and the favored amplitude  can alter the relation
between the \CP asymmetries, $A_{CP}$, for the $\jpsi\KS$ and $\jpsi\KL$
final states. In general, a difference between $A_{CP}(\jpsi\KS)$ and
$-A_{CP}(\jpsi\KL)$ of more than a few times $10^{-3}$ requires a
wrong-flavor amplitude~\cite{ligeti}. A limit on the \CP-odd part of the
phase difference between the wrong-flavor amplitude and the favored
amplitude can be derived from the measured values of \stwob from $B$
decays to the $\jpsi\KS$ and $\jpsi\KL$ final states. No test of the
modulus of the wrong-flavor amplitude currently exists.

The decay mode $\Bz\!\to\!\jpsi\Kstarz$ proceeds via the same quark
transition as $\Bz \!\to\! \jpsi \Kz$.  The matrix elements, and therefore
the ratio of  wrong-flavor to favored amplitudes, are expected
to be similar for $\Bz\!\to\!\jpsi\Kstarz$ and $\Bz \!\to\! \jpsi
\Kz$~\cite{ligeti}. In this Letter we present a measurement of
the ratio of wrong-flavor to favored amplitude for the decay
$\Bz\!\to\!\jpsi\Kstarz$, from the time-dependent asymmetry, where we use
$\Kstarz\!\to\! K^+\pi^-$ to identify the strangeness of the final state. The data
sample consists of about 88 million  $B\Bbar$ pairs produced in
$e^{+}e^{-}$ interactions at the $\FourS$ resonance, corresponding to an
integrated luminosity of $82\invfb$, collected with the \babar\
detector~\cite{babar-detector-nim} at the \pep2\ asymmetric-energy
collider at SLAC.  

Charged particles are detected, and their momenta measured, by a
combination of a vertex tracker consisting of five layers of
double-sided silicon microstrip detectors, and a 40-layer central drift
chamber, both operating in the 1.5-T magnetic field of a superconducting
solenoid. We identify photons and electrons  using a CsI(Tl)
electromagnetic calorimeter. Further charged particle identification is
provided by the average energy loss ($dE/dx$) in the tracking devices
and by an internally reflecting ring imaging Cherenkov detector covering
the central region.  

The analysis method is similar to that of other time-dependent mixing
measurements performed at \babar~\cite{babar-stwob-prd}. We use a sample
of events ($B_{\jpsi K\pi}$) in which one neutral $B$ meson is
reconstructed in the state $\jpsi\Kstarz$ or $\jpsi\Kstarzb$. The
$\jpsi$ meson is reconstructed through its decay to $e^+e^-$ or
$\mu^+\mu^-$, and the $\Kstarz$ ($\Kstarzb$) meson through its decay to
$K^+\pi^-$ ($K^{-}\pi^{+}$). We examine each event in this sample for
evidence that the other $B$ meson decayed either as a \Bz or \Bzb
(flavor tag).    

The pseudoscalar to vector-vector decay $\Bz\!\to\!\jpsi\Kstarz(892)$ is 
described by three amplitudes $A_0$, $A_\|$, and $A_\perp$, for the
longitudinal, parallel, and perpendicular transverse
polarization~\cite{anglesprl}, respectively, of the vector mesons. In
the selection of $\Bz\!\to\!\jpsi\Kstarz(892)$ there is a small
contribution from $\Bz\!\to\! \jpsi\Kstar_{0}(1430)$, whose decay
amplitude is denoted with $A_{s}$. The favored decay amplitudes
$A_{\lambda}(\Bz\!\to\!\jpsi K^+\pi^-) = a_\lambda
e^{i\delta^a_\lambda}e^{+i\phi^a}$ are described by the magnitudes
$a_\lambda$, weak phase $\phi^{a}$, and strong phases
$\delta_\lambda^{a}$, where $\lambda=0$,$\|$,$\perp$,$s$. The amplitudes
for the wrong-flavor decays are given by $A_{\lambda}(\Bzb\!\to\!\jpsi 
K^+\pi^-) = b_\lambda e^{i\delta^b_\lambda}e^{+i\phi^b}$. The
corresponding decay amplitudes for the charge-conjugate final state
$\jpsi K^-\pi^+$ are obtained by replacing $\phi^{a}$ with
$-\bar{\phi}^{a}$, $b_\lambda$ with $\bar{b}_{\lambda}$,
$\delta_\lambda^{b}$ with $\bar{\delta}_\lambda^{b}$, and $\phi^{b}$
with $-\bar{\phi}^{b}$. We assume $a_\lambda = \bar{a}_\lambda$. 

The proper-time distributions of $B$ meson decays to $\jpsi K^+\pi^-$
($\jpsi K^-\pi^+$), having either a \Bz or \Bzb tag, can be expressed in
terms of  the \Bz-\Bzb oscillation amplitude and the amplitudes
describing \Bzb and \Bz decays to this final state~\cite{lambda}. The
angular-integrated decay rate ${\rm f}_+({\rm f}_-)$ to the final state 
$\jpsi K^+\pi^-$ when the tagging meson is a $\Bz (\Bzb)$ is given by 
\begin{eqnarray}
{\rm f}_{\pm}(\deltat) = {\frac{e^{{- \left| \deltat \right|}/\tau_{\Bz} }}{4\tau_{\Bz} }}  
\Bigg[ 1 \Bigg.& \!\!\! \mp& \!\!\!  \CKS \cos{( \Delta m_{d}  \deltat )} \nonumber \\
 &\!\!\! \pm& \!\!\! \Bigg. \SKS \sin{( \Delta m_{d}  \deltat) }  \Bigg],
\label{eq:timedist-jpsikst}
\end{eqnarray}
where $\Delta t \equiv t_{\rm rec} - t_{\rm tag}$ is the difference
between the proper decay times of the reconstructed $B$ meson ($B_{\rm
  rec}$) and the tagging $B$ meson ($B_{\rm tag}$), $\tau_{\Bz}$ is the 
\Bz lifetime, and \deltamd is the \Bz-\Bzb oscillation frequency. The
corresponding decay rates ${\rm \overline{f}}_+$ and ${\rm
  \overline{f}}_-$ for the charge-conjugate final state $\jpsi K^-\pi^+$
are obtained by replacing $\CKS$ with $-\CKSB$ and $\SKS$ with $-\SKSB$. 

The \CKS\ and \SKS\ coefficients are related to the wrong-flavor and
favored amplitudes by   
\begin{equation}\label{eq:snc}
\CKS  = \frac{a^2-b^2}{a^2+b^2},\,\,\, \mbox{and}\,\,\,
\SKS  = \frac{2 \sum_\lambda \eta\, a_\lambda b_\lambda \sin(\phi + \delta_\lambda)}{a^2 + b^2},
\end{equation}
with $a^{2}\equiv a_0^{2}+a_\|^{2}+a_\perp^{2}+a_s^2$, $b^{2}\equiv
b_0^{2}+b_\|^{2}+b_\perp^{2}+b_s^2$, and $\eta=+1\ (-1)$ for $\lambda =
0,\|,s\ (\perp)$. The strong and weak phase differences are given by 
$\delta_\lambda = \delta_\lambda^b - \delta_\lambda^a$ and $\phi =
\arg(q/p) + (\phi_b - \phi_a)$, respectively, where $(q/p)$ contains the
weak phase of \Bz-\Bzb oscillations. The $\CKSB$ and $\SKSB$
coefficients are given by the same expressions, replacing
$b_{(\lambda)}$ with $\bar{b}_{(\lambda)}$, $\delta_\lambda$ with
$\bar{\delta}_\lambda$, and $\phi$ with $-\bar{\phi}$. 

In the $B\!\to\!\jpsi\Kstarz$ selection, a \jpsi\ candidate must consist 
of two identified lepton tracks~\cite{babar-detector-nim} that form a
good vertex. The lepton-pair invariant mass must be in the range $3.06 -
3.14 \gevcc$ for muons and $2.95 - 3.14 \gevcc$ for electrons. This
corresponds to a $\pm3\sigma$ interval for muons, and, for electrons,
accommodates the remaining radiative tail after bremsstrahlung
correction~\cite{babar-stwob-prd}. We form $K^+\pi^-$ candidate pairs,
where the track that is most consistent with being a kaon is assigned to
be the kaon candidate. The $K^+\pi^-$ pair must have an invariant mass
within $100 \mevcc$ of the nominal $\Kstarz (892)$
mass~\cite{PDG2002}. In the  selected mass window the $\Kstar_{0}(1430)$
contributes $(7.3\pm 1.6)\%$ of the $K^+\pi^-$ events. 

The \B -meson candidates are formed from \jpsi\ and $K^+\pi^-$ candidates
with the requirement that the difference $\Delta E = E^{\rm cm}_B -
E^{\rm cm}_{\rm beam}$ between their energy and the beam energy in the
center-of-mass frame be less than $30~\mev$ from zero. The
beam-energy-substituted mass $\mes = \sqrt{(E_{\rm beam}^{\rm cm})^2 - (p^{\rm
cm}_B)^2}$ must be greater than 5.2~\gev/$c^2$, where $p^{\rm cm}_B$ is
the measured $B$ momentum in the center-of-mass frame. We define a
signal region with $\mes > 5.27\ \gev/c^2$ to determine event yields 
and purities, and a sideband region with $\mes < 5.27\ \gev/c^2$ to
study background properties. If several \B\ candidates are found in an 
event, the one with the smallest $|\Delta E|$ is retained.  

A measurement of the asymmetry coefficients \CKS, \SKS, \CKSB, and
\SKSB\ requires a determination of the experimental $\Delta t$
resolution and the fraction $w$ of events in which the flavor tag
assignment is incorrect. This mistag fraction reduces the amplitudes of
the observed asymmetries by a factor $1-2w$. Mistag fractions and
$\Delta t$ resolution functions are determined from a sample of neutral
$B$ mesons that decay to final states with one charmed meson ($B_{Dh}$),
and consists of the channels $D^{(*)-}h^+\ (h^+=\pi^+,\rho^+$, and
$a_1^+)$. 

The algorithm for $B$-flavor tagging is explained in
Ref.~\cite{babar-stwob-prl2}. The total efficiency for assigning a 
reconstructed $B$ candidate to one of four hierarchical, mutually
exclusive tagging categories is $(65.6\pm 0.5)\%$. Untagged events are
excluded from further consideration. The effective tagging efficiency $Q
\equiv \sum_i {\eps_i (1-2\mistag_i)^2} $, where $\eps_i$ and $\mistag_i$
are the efficiencies and mistag probabilities, for events
tagged in category $i$, is measured to be $(28.1 \pm
0.7)\%$.     

The time interval \deltat between the two $B$ decays is calculated
from the measured separation \deltaz between the decay vertices of
the $B_{\rm rec}$ and $B_{\rm tag}$ along the collision ($z$)
axis~\cite{babar-stwob-prd}. We determine the $z$ position of the
$B_{\rm rec}$ vertex from its charged tracks. The $B_{\rm tag}$ vertex 
is determined by fitting tracks not belonging to the $B_{\rm rec}$ 
candidate to a common  vertex, employing constraints from the beam spot 
location and the $B_{\rm rec}$ momentum~\cite{babar-stwob-prd}. 
We accept events with a \deltat\ uncertainty of less than 2.5\ps
and $\vert \deltat \vert <20 \ps$. The fraction of events satisfying
these requirements is 95\%. 

\begin{figure}[!h]%1
  \begin{center}%
    \epsfig{figure=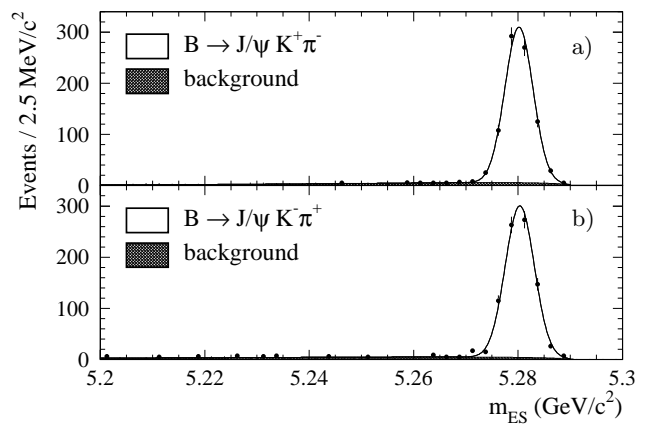,bbllx=0,bblly=0,bburx=555,bbury=355,width=0.97\linewidth}
    \put(-25,140){{a)}}
    \put(-25,75){{b)}}
    \caption{
      Distributions of \mes\ a) for $\jpsi K^+\pi^-$ candidates and b) for 
      $\jpsi K^-\pi^+$ candidates satisfying the tagging and vertexing
      requirements. The fit is described in the text.}   
    \label{fig:bcpsample}
  \end{center}
\end{figure}

Figure~\ref{fig:bcpsample} shows the \mes distributions of the $\jpsi
K^+\pi^-$ and $\jpsi K^-\pi^+$ candidates that satisfy the tagging and
vertexing requirements. The \mes distributions are fit with the sum of a
threshold function~\cite{Argus}, which accounts for the background from
random combinations of tracks in the event, and a Gaussian distribution 
describing the signal. In Table~\ref{tab:result} we list the event
yields and signal purities for the tagged $B\!\to\! \jpsi K^+\pi^-$ and
$B\!\to\! \jpsi K^-\pi^+$ candidates. The fraction of events in the
Gaussian component of the \mes fits due to other $B$ decay modes is
estimated to be $(1.6\pm0.4)\%$ based on simulated events.  

\begin{table}[!b] 
\caption{ 
Number of events, $N_{\rm tag}$, and signal purity, $P$, in the signal
region for the $\jpsi K^+\pi^-$ and $\jpsi K^-\pi^+$ samples, and for the
$B_{Dh}$ sample. Errors are statistical only.}
\label{tab:result} 
\begin{ruledtabular} 
\begin{tabular*}{\hsize}{ l@{\extracolsep{0ptplus1fil}} r c@{\extracolsep{0ptplus1fil}} D{,}{\ \pm\ }{-1} } 
 Sample  & $N_{\rm tag}$ & $P(\%)$ \\ 
\hline
$\jpsi K^+\pi^-$ sample   & $860$    & $95.5\pm 0.7$ \\%
$\jpsi K^-\pi^+$ sample   & $856$    & $96.5\pm 0.6$ \\%
\hline
$B_{Dh}$ sample           & $25375$  & $84.9\pm 0.2$      \\ 
\end{tabular*} 
\end{ruledtabular} 
\end{table}

We determine the \CKS , \SKS , \CKSB , and \SKSB\ coefficients with a
simultaneous unbinned maximum likelihood fit to the \deltat
distributions of the tagged $B_{\jpsi K\pi}$ and $B_{Dh}$ samples. In
this fit the \deltat\ distributions of the $\jpsi K^+\pi^-$ and 
$\jpsi K^-\pi^+$ samples are described by Eq.~(\ref{eq:timedist-jpsikst}).
The \deltat distributions of the $B_{Dh}$ sample are described by
the same equation with $C=1$ and $S=0$. The observed amplitudes for the
time-dependent asymmetries in the $B_{\jpsi K\pi}$ sample and for 
flavor oscillation in the $B_{Dh}$ sample are reduced by the same
factor, $1-2\mistag$, due to flavor mistags. Events are 
assigned signal and background probabilities based on the \mes\
distributions. The \deltat distributions for the signal are convolved
with a common resolution function, modeled by the sum of three
Gaussians~\cite{babar-stwob-prd}. Backgrounds are incorporated by means
of an empirical description of their \deltat spectra, 
obtained from the \mes-sideband region,
containing prompt and  non-prompt components convolved with a resolution
function~\cite{babar-stwob-prd} distinct from that of the signal.

There are 48 free parameters in the fit. The fit parameters that
describe the signal \deltat distributions are \CKS , \SKS , \CKSB , and
\SKSB\ (4), the average mistag fraction $\mistag$, the difference
$\Delta\mistag$ between \Bz\ and \Bzb\ mistag fractions, and the linear
dependence of the mistag fraction on the \deltat error for each tagging
category (12), parameters for the signal \deltat resolution (8), and
parameters to account for differences in reconstruction and tagging
efficiencies for \Bz\ and \Bzb mesons (5). The $B_{\jpsi K\pi}$ and
$B_{Dh}$ background \deltat distributions are described by parameters
for the background time dependence (8),  \deltat resolution (3), and
mistag fractions (8). We fix $\tau_{\Bz}$ at $1.542\ps$ and $\deltamd$
at $0.489\ps^{-1}$~\cite{PDG2002}. The determination of the mistag
fractions and \deltat resolution function parameters for the signal is
dominated by the large $B_{Dh}$ sample.  Background parameters are
determined from events with $\mes < 5.27\gevcc$. 

\begin{figure}[ht]
\begin{center}
\epsfig{figure=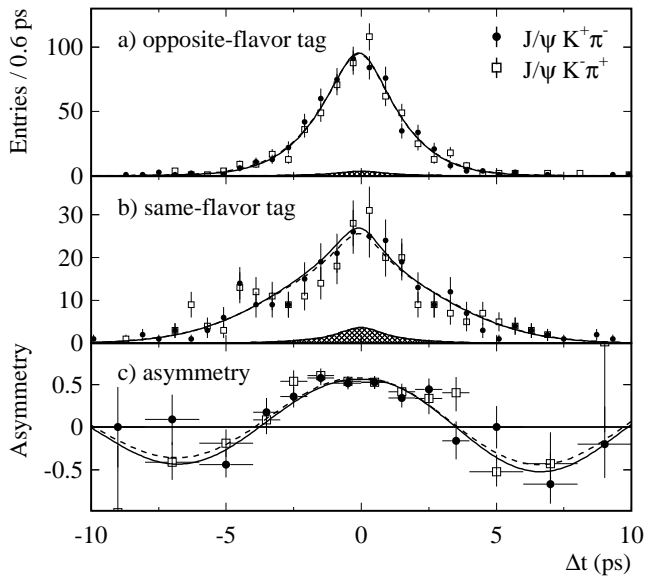,bbllx=126,bblly=20,bburx=457,bbury=325,width=0.97\linewidth} 
\caption{
Number of $\jpsi K^+\pi^-$ and $\jpsi K^-\pi^+$
candidates in the signal region 
a) with an opposite-flavor $B$ tag, $N_{OF}$, 
b) with a same-flavor $B$ tag, $N_{SF}$, and 
c) the observed asymmetry $(N_{OF}-N_{SF})/(N_{OF}+N_{SF})$ as functions of 
\deltat. In each figure the solid (dashed) curves represent the fit
projection in \deltat for $\jpsi K^+\pi^-$ $(\jpsi K^-\pi^+)$
candidates. The shaded regions in (a) and (b) represent the background
contributions.} 
\label{fig:cpdeltat}
\end{center}
\end{figure}

The fit to the $B_{\jpsi K\pi}$ and $B_{Dh}$ samples yields
$\CKS  = 1.045  \pm 0.058 \pm 0.035$,
$\SKS  = -0.024 \pm 0.095 \pm 0.041$,
$\CKSB = 0.966  \pm 0.051 \pm 0.035$, and
$\SKSB = 0.004  \pm 0.090 \pm 0.041$,
where the first error is statistical and the second error is
systematic. Figure~\ref{fig:cpdeltat} shows the \deltat distributions
and the asymmetries in yields between \Bz tags and \Bzb tags as a
function of \deltat for the $\jpsi K^+\pi^-$ and $\jpsi K^-\pi^+$
samples, overlaid with the projection of the likelihood fit result. 

We estimate common systematic errors for \CKS\ (\SKS) and \CKSB\
(\SKSB). The dominant sources of systematic error are the uncertainties 
in the level, composition, and time-dependent asymmetry of the
background in the selected $B_{\jpsi K\pi}$ sample (0.016 for \CKS,
0.017 for \SKS), uncertainties in the beam spot location and the
internal alignment of the vertex detector (0.016 for \CKS, 0.021 for
\SKS), and the statistics of the simulated event sample (0.016 for \CKS,
0.015 for \SKS). Another significant contribution to the systematic
uncertainty in the cosine coefficients comes from possible differences
between the $B_{Dh}$ and $B_{\jpsi K\pi}$ mistag fractions (0.012). The
uncertainty in the interference between the suppressed $\bar b\!\to\! 
\bar u c \bar d$ amplitude with the favored $b\!\to\! c \bar u d$
amplitude for the decay modes in the $B_{Dh}$ sample and for certain
tag-side $B$ decays to hadronic final states~\cite{dcsd} contributes to
the systematic uncertainty in the sine coefficients (0.019). Finally,
there are differences in the angular-integrated efficiency for the
$B\!\to\!\jpsi\Kstarz(892)$ helicity amplitudes and  the
$B\!\to\!\jpsi\Kstar_{0}(1430)$ amplitude (0.007 for \CKS, 0.016 for
\SKS). The total systematic errors for the cosine coefficients and sine 
coefficients are $0.035$ and $0.041$, respectively. Most systematic
errors are determined with data and are expected to decrease with
larger sample size.    

The large $\jpsi K^+\pi^-$ and $\jpsi K^-\pi^+$ samples allow a number
of consistency checks, including separation by data-taking period and
tagging category. The results of fits to these subsamples are found to
be statistically consistent.  

The  measured values of the cosine and sine coefficients are consistent
with $\CKS=\CKSB =1$ and $\SKS = \SKSB =0$, as expected for no
contributions from the wrong-flavor decays $\Bz\!\to\!\jpsi K^-\pi^+$
and $\Bzb\!\to\!\jpsi K^+\pi^-$. We use the measured cosine coefficients
\CKS\ and \CKSB\ and assume $|q/p|=1$~\cite{babar-dilepton} to calculate
the wrong-flavor to favored decay rate ratios $\Gamma (\Bzb\!\to\!\jpsi
K^+\pi^-)/\Gamma (\Bz\!\to\!\jpsi K^+\pi^-) = |b/a|^2=-0.022 \pm 0.028\
\mbox{(stat.)} \pm 0.016\ \mbox{(syst.)}$ and $\Gamma (\Bz\!\to\!\jpsi
K^-\pi^+)/\Gamma (\Bzb\!\to\!\jpsi K^-\pi^+) = |\bar{b}/a|^2=0.017 \pm
0.026\ \mbox{(stat.)}  \pm 0.016\ \mbox{(syst.)}$, where the negative
central value occurs because $C>1$. From these measurements the
wrong-flavor to favored amplitude ratios for $B\!\to\!
\jpsi\Kstarz(892)$ and $B\!\to\! \jpsi\Kstarzb(892)$ can be
calculated. Using the measured fraction of $B\!\to\!
\jpsi\Kstar_{0}(1430)$ events contributing in the $B\!\to\! \jpsi 
K^+\pi^-$ selection, the upper limits for the decay amplitude ratios at
$90\%$ confidence level (C.L.) are found to be $|A
(\Bzb\!\to\!\jpsi\Kstarz)|/|A (\Bz\!\to\!\jpsi\Kstarz)| < 0.26$ and $|A
(\Bz\!\to\!\jpsi\Kstarzb)|/|A (\Bzb\!\to\!\jpsi\Kstarzb)| < 0.32$. For
the single ratio of wrong-flavor to favored amplitude for \Bz and \Bzb
combined, we determine an upper limit of $0.25$ at $90\%$ C.L.

In conclusion, we observe no evidence for the wrong-flavor decays
$\Bzb\!\to\!\jpsi\Kstarz(892)$ and
$\Bz\!\to\!\jpsi\Kstarzb(892)$. Together with theoretical information on
the relation between the matrix elements for $\Bz\!\to\! \jpsi K^0$ and
$\Bz\!\to\!\jpsi K^{*0}$~\cite{ligeti}, the results presented here can
be used to set a limit on the difference between $A_{CP}(\jpsi\KS)$ and 
$-A_{CP}(\jpsi\KL)$.  

We are grateful for the excellent luminosity and machine conditions
provided by our \pep2\ colleagues, 
and for the substantial dedicated effort from
the computing organizations that support \babar.
The collaborating institutions wish to thank 
SLAC for its support and kind hospitality. 
This work is supported by
DOE
and NSF (USA),
NSERC (Canada),
IHEP (China),
CEA and
CNRS-IN2P3
(France),
BMBF and DFG
(Germany),
INFN (Italy),
FOM (The Netherlands),
NFR (Norway),
MIST (Russia), and
PPARC (United Kingdom). 
Individuals have received support from the 
A.~P.~Sloan Foundation, 
Research Corporation,
and Alexander von Humboldt Foundation.


\begin{thebibliography}{99}
\bibitem{CKM}
\hyphenation{Ko-ba-ya-shi}
N.~Cabibbo, \prl {\bf 10}, 531 (1963); 
M.~Kobayashi and T.~Maskawa, Prog.\ Theor.\ Phys.\ {\bf 49}, 652
(1973). 

\bibitem{BCP}
A.B.~Carter and A.I.~Sanda, \prd {\bf 23}, 1567 (1981);
I.I.~Bigi   and A.I.~Sanda, \np {\bf B193}, 85 (1981).

\bibitem{ligeti}
Y.~Grossman, A.L.~Kagan, and Z.~Ligeti, 
\plb {\bf 538}, 327 (2002).

\bibitem{chargeconj}
Charge conjugation is implied throughout this letter, unless explicitly 
stated otherwise. 

\bibitem{babar-detector-nim}
\babar\ Collaboration, B.\ Aubert {\em et al.}, Nucl.\ Instrum.\ Methods
Phys.\ Res., Sect.\ A {\bf 479}, 1 (2002).

\bibitem{babar-stwob-prd}
\babar\ Collaboration, B.\ Aubert {\em et al.}, 
\prd {\bf 66}, 032003 (2002).

\bibitem{anglesprl}
A.S.~Dighe, I.~Dunietz, H.J.~Lipkin, and J.L.~Rosner,
\plb {\bf 369}, 144 (1996).

\bibitem{lambda}
See, for example, L.~Wolfenstein, in \prd {\bf 66}, 010001 (2002).

\bibitem{PDG2002}
Particle Data Group, K.~Hagiwara {\em et al.}, \prd {\bf 66}, 010001
(2002). 

\bibitem{babar-stwob-prl2}
\babar\ Collaboration, B.\ Aubert {\em et al.},
\prl {\bf 89}, 201802 (2002).

\bibitem{Argus}
ARGUS Collaboration, H.~Albrecht {\em et al.}, \zpc{\bf 48}, 543
(1990). 

\bibitem{dcsd}
O.~Long, M.~Baak, R.N.~Cahn, and D.~Kirkby, \prd {\bf 68}, 034010
(2003). 

\bibitem{babar-dilepton}
\babar\ Collaboration, B.\ Aubert {\em et al.}, 
\prl {\bf 88}, 231801 (2002).
\end{thebibliography}
\end{document}